\documentclass[superscriptaddress,twocolumn,nofootinbib]{revtex4}
\usepackage{amssymb}
\usepackage{amsfonts}
\usepackage{amsmath}
\usepackage{amsthm}
\usepackage{graphicx}
\usepackage{dcolumn}
\usepackage{bm}
\usepackage[english]{babel}
\usepackage{enumerate}
\usepackage[T1]{fontenc}

\begin{document}

\title{A New Approach to Frequency Independent Radiating Systems: Conformal Edge Antennas}

\author{Franco Fiorini}
\email{francof@cab.cnea.gov.ar.}
\affiliation{Departamento de Ingeniería en Telecomunicaciones and Instituto Balseiro, Centro Atómico Bariloche, Av. Ezequiel Bustillo 9500, CP8400, S.C. de Bariloche, Río Negro, Argentina.}
\author{Leonardo Morbidel}
\email{leomorbidel@ib.edu.ar}
\affiliation{Departamento de Ingeniería en Telecomunicaciones and Instituto Balseiro, Centro Atómico Bariloche, Av. Ezequiel Bustillo 9500, CP8400, S.C. de Bariloche, Río Negro, Argentina.}
\author{Pablo Costanzo Caso}
\email{pcostanzo@ib.edu.ar}
\affiliation{Departamento de Ingeniería en Telecomunicaciones and Instituto Balseiro, Centro Atómico Bariloche, Av. Ezequiel Bustillo 9500, CP8400, S.C. de Bariloche, Río Negro, Argentina.}


\begin{abstract}
Conformal techniques, based on the covariance of Maxwell's electrodynamics under the full conformal group in four spacetime dimensions, are discussed in relation with the constant input impedance properties of frequency-independent antennas. In particular we show that by applying suitable conformal transformations to existing self-complementary planar structures like logarithmic spirals, the constant input impedance property is considerably improved, specially in the low frequency domain, where it usually fails due to the large wavelength involved. The effect of the conformal transformation is to bring infinity up to a finite distance and, in a way, to imprison the otherwise infinite structure in a compact region. This procedure enables to perform a more intelligent truncation with the aim to capture some of the properties left behind in the cutting, and to design realistic ultra width-band antennas with a more controlled value of the input impedance in the whole operation range. Due to the fact that the tools here developed find their roots in the differential geometry of curved manifolds, they are quite general and also suitable for dealing with 3D-antennas of arbitrary shape, curvature and size, as well as with other systems in which the effects of truncation play a crucial role in defining physical properties of them.
\end{abstract}

\maketitle

\section{Introduction}

Ultra wide band (UWB) technology has been of considerable interest in the scientific community since the usable frequency bands and the maximum radiated power within them, were defined. One of the main advantages of UWB systems is the reduction of the fading channel because their narrow band response. Although the transmission rate is reduced due to multi-path, sensing and localization applications are greatly benefited \cite{UWB1}. Currently, many of the investigations in UWB technology are oriented to medical purposes, in particular, to the detection of inhomogeneities in the body, such as tumors \cite{UWB1b}. Recently explored fields for UWB are, for instance, the \emph{body area networks} for medical surveying and data transmission \cite{UWB1c}, as well as the development of photonic devices for UWB signal processing \cite{UWB1d}. In all UWB applications the development of broad band antennas continues to be a key and critical point, which is intensively investigated nowadays \cite{UWB2}-\cite{UWB5}.

The concept of \emph{frequency-independent antennas} (FIA's) was formally introduced by Rumsey long time ago \cite{Rumsey}, even though several investigations on this topic can be traced back to the early work of Mushiake \cite{Mushiake1}, specially the ones concerning the constant input impedance of \emph{self-complementary} (SC) or \emph{mutually dual} structures, whose radiative properties are still a matter of an active research \cite{Nipones}. Currently, FIA's are an indivisible part of UWB wireless communication systems, and their use is widely extended in practically any of the applications mentioned above.

Basically, when one refers to frequency-independent or UWB antennas, one is thinking about those in which some of their main characteristics (as the input impedance, the radiation pattern, the gain, the polarization, among others), remain reasonably constant over a wide range of frequencies. For instance, and depending on the application in mind, one of the key aspects of UWB antenna design could lie in an appropriate control of the input impedance along the whole operation bandwidth, which usually will cover several GHz at microwave bands. If this were the case, SC antennas would be of great help by virtue of the constant input impedance predicted by Mushiake's relation, which emerges out by analyzing the radiative properties of SC structures. However, this relation is based on the assumption that both the antenna and its twin dual structure are of infinite size, and it constitutes just an approximation in actual radiating systems, which are subjected to truncation processes of some sort. Typical deterioration of the constant input impedance property comes from the fact that there will be some harmful effects at the feeding point as a result of reflected currents from the truncated ends, and from the distortion in the structure introduced by the feeding port itself.

In this paper we shall first expose the rudiments concerning the conformal invariance of $D=4$ Maxwell's electrodynamics, established more than a century ago by Cunningham and Bateman \cite{Bateman1}-\cite{Bateman}, and its subsequent application in regard with the property of constant input impedance of self complementary radiating systems. Even though the nature of the conformal group was well understood in mathematical physics since the beginning of the twentieth century, and that its importance in modern physics --specially in quantum field theory and string theory-- is very well established \cite{Fulton}-\cite{Kastrup}, its connection with antenna theory, to say the least, was barely reported in the literature. The purpose of section \ref{secciondos} is to give a brief account on the matter, not only with the aim of introducing the basic concepts and notation related to the geometry of Minkowski spacetime and the conformal group acting on it, but also to expose the underlying ideas in a more comprehensive and concise manner to the electric engineer interested in antenna theory and design, to the physicist whose domain is applied physics, and to the mathematician which enjoys seeing how the science he/she profess is applicable to concrete, engineering problems.

After reviewing the main properties and concepts surrounding FIA's in section \ref{secciontres}, we expose in section \ref{seccioncuatro} the central idea of the paper. By means of conformal techniques used frequently in the analysis of the large scale properties of space-times, we propose a novel antenna truncation method which, in an appropriated mathematical sense, takes into account in the design the distant regions of space which were simply discarded in the usual approach. This allows to construct more compact antennas with better input impedance performance, in the sense that the constant impedance characteristic of SC structures is optimized. Several \emph{conformal edge spirals antennas} (the concept here introduced), based on standard logarithmic spirals, are simulated and their input impedances are compared in section \ref{seccioncinco}. We finally expose our conclusions as well as some future lines of research in section \ref{conclusiones}.

\section{Conformal invariance of Electrodynamics}\label{secciondos}

\subsection{A brief passage through the conformal group}

We start reviewing some key concepts associated to the conformal group that have a direct implication for our work. For details and a complete exposition we direct the reader to the standard references in the field, e.g., \cite{Conformal}.
Even though we will work in a flat (i.e., non gravitational) background, it is more natural to deal with conformal symmetry in a general spacetime $(\mathcal{M},\textbf{g})$, where $\mathcal{M}$ is a pseudo-Riemannian (Lorentzian) $D$-dimensional manifold ($n\geq2$), and $\textbf{g}$ a Lorentzian metric which allow us to measure distances on $\mathcal{M}$ in the way\footnote{We shall adopt throughout the work the Einstein's summation convention.}
\begin{equation}
ds^{2}=g_{\mu\nu}(x)dx^{\mu}dx^{\nu},\label{metrica}
\end{equation}%
being $g_{\mu\nu}(x)$ the local components of $\textbf{g}$ in a local coordinate system $x^{\mu}$. This metric enables us to know the length of a vector, as well as the angle between two vectors in a certain point $x$ in $\mathcal{M}$. If $V$ and $W$ are two vectors in $T_{x}\mathcal{M}$ (the tangent space of $\mathcal{M}$ at the point $x$), we have in local coordinates that
\begin{equation}
\textbf{V}=v^{\mu}\partial_{\mu},\,\,\,\,\textbf{W}=w^{\nu}\partial_{\nu}, \label{vectors}
\end{equation}%
where $\partial_{\mu}\equiv\partial_{x^{\mu}}$ are the local coordinate basis vectors. In particular we have the squared norm of the vectors at $x$ given by, for instance, $\|\textbf{V}\|^{2}=g_{\mu\nu}v^{\nu}v^{\mu}=v_{\mu}v^{\mu}$, whereas the cosine of the angle between $\textbf{V}$ and $\textbf{W}$ at the same point is expressed as

\begin{equation}
Cos(\textbf{V},\textbf{W})=\frac{g_{\mu\nu}v^{\mu}w^{\mu}}{\|\textbf{V}\|\|\textbf{W}\|}\label{cosvec}.
\end{equation}%

If we consider another $D$-dimensional spacetime $(\bar{\mathcal{M}},\bar{\textbf{g}})$ with local coordinates $\bar{x}^{\mu}$, then, a mapping $x^{\mu} \rightarrow \bar{x}^{\mu}$ is said \emph{conformal} if induces a transformation of the metric tensor of the sort
\begin{equation}
\bar{g}_{\mu\nu}(\bar{x})=\Omega(x)g_{\mu\nu}(x)\label{mapconf},
\end{equation}%
for a positive function $\Omega(x)$\footnote{We adopt here the common attitude in physics of considering $\Omega(x)$ positive in order to prevent causality violations.}. On the other hand, under a general coordinate change, the covariant components of the metric tensor transform according to

\begin{equation}
g_{\mu\nu}(\bar{x})=\frac{\partial x^{\alpha}}{\partial \bar{x}^{\mu}}\frac{\partial x^{\beta}}{\partial \bar{x}^{\nu}}g_{\alpha\beta}(x)\label{cambiomet},
\end{equation}%
so, the Eq. (\ref{mapconf}) just says

\begin{equation}
\bar{g}_{\mu\nu}(\bar{x})=\Omega(x) \frac{\partial x^{\alpha}}{\partial \bar{x}^{\mu}}\frac{\partial x^{\beta}}{\partial \bar{x}^{\nu}}g_{\alpha\beta}(x).\label{concambio}
\end{equation}%

Equation (\ref{cosvec}) suggests why the conformal mappings bear their name; the angle between two arbitrary curves crossing each other at some point does not change by the effect of conformal scalings of the metric tensor.

The set of conformal transformations forms a Lie group, and in the context alluded in Eq. (\ref{mapconf}), it obviously includes the Poincare group as a subgroup, since the latter corresponds to set $\Omega(x)=1$. In order to take contact with the group generators, we shall consider infinitesimal coordinate changes of the form $x^{\mu}\rightarrow \bar{x}^{\mu}=x^{\mu}+\xi^{\mu}(x)$. If we keep only first order terms in the infinitesimal generator $\xi^{\mu}(x)$, Eq. (\ref{cambiomet}) reduces to
\begin{equation}
g_{\mu\nu}(\bar{x})=g_{\mu\nu}(x)-(\partial_{\mu}\xi_{\nu}(x)+\partial_{\nu}\xi_{\mu}(x))  \label{cambiometinfi}.
\end{equation}%
In order for the transformation to be conformal, it is necessary that
\begin{equation}
\partial_{\mu}\xi_{\nu}(x)+\partial_{\nu}\xi_{\mu}(x)=\sigma(x) g_{\mu\nu},  \label{condicion}
\end{equation}%
for a function $\sigma(x)$ such that $\Omega=1-\sigma(x)>0$, but otherwise, arbitrary. By contracting expression (\ref{condicion}) with the inverse metric $g^{\mu\nu}$, we can determine the function $\sigma(x)$ in terms of the generator $\xi^{\mu}(x)$, i.e.
\begin{equation}
\sigma(x)=2D^{-1}\partial_{\mu}\xi^{\mu}(x),  \label{traza}
\end{equation}%
where we have used $g^{\mu\nu}g_{\mu\nu}=D$. Even though the present treatment applies to any space-time $(\mathcal{M},\textbf{g})$, we will work in a flat Minkowskian background with its canonical metric tensor $\eta_{\mu\nu}=diag(-1,1,...,1)$ of Lorentz signature $D-2$. In order to proceed, we will take first an additional derivative $\partial_{\rho}$ in expression (\ref{condicion}), permute cyclically the indices $\{\rho\mu\nu\}$ on it latter, and finally take the linear combination $-\{\rho\mu\nu\}+\{\nu\rho\mu\}+\{\mu\nu\rho\}$, obtaining
\begin{equation}
2\partial_{\mu}\partial_{\nu}\xi_{\rho}=\eta_{\rho\mu}\partial_{\nu}\sigma+\eta_{\rho\nu}\partial_{\mu}\sigma-\eta_{\nu\mu}\partial_{\rho}\sigma,
 \label{permutacion}
\end{equation}%
where the argument in $\xi_{\rho}(x)$ and $\sigma(x)$ should be understood hereafter. Upon contracting this last equation with $\eta^{\mu\nu}$, we get

\begin{equation}
2\partial^{2}\xi_{\mu}=(2-D)\partial_{\mu}\sigma,
 \label{permuCON}
\end{equation}%
where $\partial^{2}\equiv\partial^{\mu}\partial_{\mu}$, and we have used several times that $\eta_{\mu\nu}\eta^{\nu\rho}=\delta_{\mu}^{\rho}$. As one might intuit, expression (\ref{permuCON}) heralds the special role played by the conformal mappings in $D=2$. Finally, we can apply $\partial_{\nu}$ to (\ref{permutacion}) and $\partial^{2}$ to (\ref{condicion}) and to combine the results to obtain
\begin{equation}
\eta_{\mu\nu}\partial^{2}\sigma=(2-D)\partial_{\mu}\partial_{\nu}\sigma,
 \label{permuCON2}
\end{equation}%
which leads, after taking trace once more, to
\begin{equation}
(D-1)\partial^{2}\sigma=0.
 \label{permufin}
\end{equation}%
This last result expresses the fact that the conformal symmetry is trivial in $D=1$, because any smooth local 1-dimensional transformation is conformal.

The results just found enable us to fully characterize the function $\sigma(x)$. Actually, if $D>2$ (the case $D=2$ will be synoptically developed below), Eqs. (\ref{permuCON2}) and (\ref{permufin}) tell us that $\partial_{\mu}\partial_{\nu}\sigma=0$, and then $\sigma(x)=A+B_{\mu}x^{\mu}$, with constants $A$ and $B_{\mu}$. In turn, by means of Eq. (\ref{permutacion}), this implies that $\partial_{\mu}\partial_{\nu}\xi_{\rho}(x)$ is constant, so we conclude that
\begin{equation}
\xi_{\mu}(x)=a_{\mu}+b_{\mu\nu}x^{\nu}+c_{\mu\nu\rho}x^{\nu}x^{\rho},
 \label{geninf}
\end{equation}%
with constants $a_{\mu},b_{\mu\nu},c_{\mu\nu\rho}$ and $c_{\mu\nu\rho}=c_{\mu\rho\nu}$. Having obtained the functional form of the generators, we shall find the physical meaning of the parameters $a_{\mu},b_{\mu\nu},c_{\mu\nu\rho}$. In order to do that, we substitute (\ref{geninf}) in (\ref{condicion}), and proceed to solve order by order in $x^{\nu}$. The constant term $a_{\mu}$ imposes no constraint at all, because both sides of the equation are just null (use Eq. (\ref{traza}) in the right hand side of (\ref{condicion})). This constant term is representative of the four translations given by $\bar{x}^{\mu}=x^{\mu}+a^{\mu}$.

The linear term $b_{\mu\nu}$ give us
\begin{equation}
b_{\mu\nu}+b_{\nu\mu}=2 D^{-1}b^{\lambda}_{\,\,\lambda}\,\eta_{\mu\nu}.
 \label{terlin}
\end{equation}%
The structure of this equation lead us to
\begin{equation}
b_{\mu\nu}=\epsilon\, \eta_{\mu\nu}+ \omega_{\mu\nu},
 \label{descom}
\end{equation}%
 with free parameters $\epsilon$ and $\omega_{\mu\nu}=-\omega_{\nu\mu}$. This is no more than the decomposition of $b_{\mu\nu}$ in its symmetric and skew-symmetric parts, being the symmetric part, a pure trace term. The constant $\epsilon$ represents an infinitesimal scale transformation of the coordinates, while $\omega_{\mu\nu}$ are the infinitesimal generators of the Lorentz group $SO(D-1,1)$, i.e., they represent the $D-1$ \emph{boosts} and $(D-1)(D-2)/2$ \emph{rotations} which leave invariant the Minkowski metric $\eta_{\mu\nu}$. The so called \emph{special conformal transformations} (SCT's), which actually change the metric by generating an $\Omega(x)\neq1$ conformal factor, will arise in considering the effect of the quadratic term $c_{\mu\nu\rho}$ in (\ref{condicion}). After permuting indices and combining in the manner we did before, it is not hard to obtain
\begin{equation}
c_{\mu\nu\rho}=b_{\nu}\,\eta_{\mu\rho}+b_{\rho}\,\eta_{\mu\nu}-b_{\mu}\,\eta_{\nu\rho},\,\,\,\,\,b_{\mu}\equiv D^{-1}\,c^{\lambda\lambda}_{\,\,\,\,\,\,\,\mu}. \label{tercuad}
\end{equation}%
 This expression for $c_{\mu\nu\rho}$ lead us to the infinitesimal special conformal transformation
 \begin{equation}
\bar{x}^{\mu}= x^{\mu} +(\textbf{b}\cdot\textbf{x}) \,x^{\mu}-b^{\mu}\,\textbf{x}^{2}, \label{scts}
\end{equation}%
where $\textbf{b}\cdot\textbf{x}=b^{\mu}x^{\nu}\eta_{\mu\nu}$ and $\textbf{x}^{2}=x^{\mu}x^{\nu}\eta_{\mu\nu}$. According to the results just obtained, we can count the number of generators of conformal transformations in $D$ spacetime dimensions; they consist of $D$ translations generated by $a^{\mu}$, one scale transformation generated by $\epsilon$ (usually called \emph{dilation}), $D(D-1)/2$ Lorentz transformations generated by $\omega_{\mu\nu}$ in (\ref{descom}), and finally, $D$ more SCT's generated by $b^{\mu}$. This give us a total of $(D^{2}+3D+2)/2$ generators, which reduces to fifteen in the $3+1$ dimensional (physical) case.

However, just the SCT's are of importance in the present work, because they are the ones that actually change the metric up to non trivial conformal factor. Regarding this, it can be seen that the most general SCT can be obtained starting from the transformation of \emph{inversion} in the hypersphere, i.e., from the mapping $x^{\mu}\rightarrow x^{'\mu}=x^{\mu}/\textbf{x}^{2}$. In fact, after performing the sequence inversion-translation-inversion given by the mappings
\begin{eqnarray}
x^{\mu}\rightarrow x^{'\mu}&=&\frac{x^{\mu}}{\textbf{x}^{2}},\notag \\
x^{'\mu}\rightarrow x^{''\mu}&=&x^{'\mu}-b^{\mu},\notag\\
x^{''\mu}\rightarrow \bar{x}^{\mu}&=&\frac{x^{''\mu}}{\textbf{x}^{''2}}, \label{secuencia}
\end{eqnarray}%
then we see that
\begin{equation}
\bar{x}^{\mu}= \frac{x^{\mu}-b^{\mu}\,\textbf{x}^{2}}{1-2\,\textbf{b}\cdot\textbf{x}+ \textbf{b}^{2}\textbf{x}^{2}}. \label{transfinita}
\end{equation}%
This is the finite version of the infinitesimal expression (\ref{scts}). It follows from (\ref{transfinita}) that
\begin{equation}
\bar{\textbf{x}}^{2}=\frac{\textbf{x}^{2}}{1-2\,\textbf{b}\cdot\textbf{x}+ \textbf{b}^{2}\textbf{x}^{2}}\equiv\frac{\textbf{x}^{2}}{\lambda(x)}, \label{cuadra}
\end{equation}%
where we have defined the function $\lambda(x)$. Then we have
\begin{equation}
(\bar{\textbf{x}}-\bar{\textbf{y}})^{2}=\lambda^{-1}(\textbf{x})\lambda^{-1}(\textbf{y})(\textbf{x}-\textbf{y})^{2}, \label{difcuadra}
\end{equation}%
so the line element reads
\begin{equation}
\bar{ds}^{2}=\lambda^{-2}(\textbf{x})ds^{2}. \label{difcuadra}
\end{equation}%
With the help of (\ref{mapconf}) we can take contact with the function $\Omega(x)$, obtaining
\begin{equation}
\Omega(x)\equiv \lambda^{-2}(\textbf{x})=\Big(1-2\,\textbf{b}\cdot\textbf{x}+ \textbf{b}^{2}\textbf{x}^{2}\Big)^{-2}. \label{twofunc}
\end{equation}%
We see that, if the vector $\textbf{b}$ is real, then naturally $\Omega(x)>0$, and the causal character of a given vector field is preserved by the conformal mapping.

As mentioned before, the case $D=2$ deserves special attention. Let us comment on this peculiar situation, and in order to get closer to the Euclidean world, let us fix $g_{\mu\nu}=\delta_{\mu\nu}$ in (\ref{condicion}), and consider coordinates $x\equiv(x^{1},x^{2})$. Combining Eqs. (\ref{condicion}) and (\ref{traza}), we immediately note that

\begin{equation}
\partial_{1}\xi_{1}(x)=\partial_{2}\xi_{2}(x),\,\,\,\,\,\partial_{1}\xi_{2}(x)=-\partial_{2}\xi_{1}(x), \label{condCR}
\end{equation}%
which are easily recognizable as the Cauchy-Riemann equations for the holomorphic (anti-holomorphic) functions\footnote{Here, $\bar{z}$ refers to the complex conjugate of $z$, and not to conformal coordinates, as in $\bar{\textbf{x}}$.} $\xi(z)=\xi^{1}(z)+ i\,\xi^{2}(z)$ ($\bar{\xi}(\bar{z})=\xi^{1}(\bar{z})-i\,\xi^{2}(\bar{z})$) of the complex variables $z,\bar{z}=x^{1}\pm\, i x^{2}$. In other words, conformal mappings in $D=2$ Euclidean plane can be interpreted as analytic coordinate transformations on the complex plane onto itself. These are of the sort $z\rightarrow f(z)$ and $\bar{z}\rightarrow \bar{f}(\bar{z})$ for some analytic function $f(z)$. In complex coordinates, the line element of $\mathbb{R}^{2}$ will transforms as

\begin{equation}
ds^{2}=dzd\bar{z}\,\rightarrow \Big|\frac{\partial f}{\partial z}\Big|^{2}\,dzd\bar{z}, \label{metrica compleja}
\end{equation}%
so the conformal factor is just $\Omega=\Big|\frac{\partial f}{\partial z}\Big|^{2}$. This kind of conformal transformation was extensively used in the past in many different contexts, such as cartography, fluid dynamics, electrodynamics, and electrical engineering in general.

\subsection{Invariance of 4D-electrodynamics under the conformal group}

Conformal invariance would be only a mathematical curiosity if physics would not be conformaly invariant. Actually, this is the case for most of modern physical theories, but fortunate, there is an exception: classical electrodynamics in $D=4$. The behavior under conformal changes of $\textbf{E}$ and $\textbf{H}$, as well as other physical quantities such the charge and current densities, was studied first by Cunningham and Bateman more than a century ago in their seminal works \cite{Cuni}, \cite{Bateman}, based on previous results from Bateman \cite{Bateman1} (for a more modern discussion we can refer the reader to \cite{Osborn}). In the following, we review the conformal invariance of $D=4$ Maxwell electrodynamics\footnote{We will take, then, $D=4$ all along this section.}.

We shall first investigate how the tensor fields transform under the conformal group, and how these transformations influence the behavior of fields and currents in electrodynamics. As mentioned before Eq. $(\ref{terlin})$, we have that the conformal group contains the translations $\bar{x}^{\mu}=x^{\mu}+a^{\mu}$ as a proper subgroup. This means that the operator $\partial_{\mu}$, which is basically the generator of translations, transforms under conformal transformations as a covariant vector, this is
\begin{equation}
\bar{\partial}_{\mu}=\frac{\partial x^{\rho}}{\partial \bar{x}^{\mu}}\partial_{\rho}.\label{transfveccor}
\end{equation}%
Note, however, that the operator $\partial^{\mu}=g^{\mu\nu}\partial_{\nu}$ do not transforms as a contravariant vector, but as a \emph{contravariant vector density}, because
\begin{equation}
\bar{g}^{\mu\nu}(\bar{x})=\Omega^{-1}(x) \frac{\partial \bar{x}^{\mu}}{\partial x^{\alpha}}\frac{\partial \bar{x}^{\nu}}{\partial x^{\beta}}g^{\alpha\beta}(x),\label{concambiocontra}
\end{equation}%
by virtue of expression (\ref{concambio}). This means that
\begin{equation}
\bar{\partial}^{\mu}=\Omega^{-1}(x) \frac{\partial \bar{x}^{\mu}}{\partial x^{\rho}}\partial^{\rho}.\label{transfveccor}
\end{equation}%
The different role played by covariant and contravariant components of a given tensor under the action of the conformal group, is a signature of it; this is due to the fact that $g^{\alpha\beta}$ transforms as a tensor density (Eq. (\ref{concambiocontra})).

In order to describe the transformation law of any field (not just $\partial_{\mu}$ or $\partial^{\mu}$), we proceed to invert the coordinates changes in (\ref{concambio}), namely
\begin{equation}
\bar{g}_{\mu\nu} \frac{\partial \bar{x}^{\mu}}{\partial x^{\alpha}}\frac{\partial \bar{x}^{\nu}}{\partial x^{\alpha}}=\Omega(x) g_{\alpha\beta}= \lambda^{-2}(x)g_{\alpha\beta}.\label{concaminv}
\end{equation}%
If we take determinant in both sides of $(\ref{concaminv})$ we obtain
\begin{equation}
\det \Big(\frac{\partial \bar{x}}{\partial x}\Big)=\lambda^{-4}(x)\label{det}.
\end{equation}%
In view of this, we can introduce the following matrix
\begin{equation}
\Lambda^{\mu}_{\,\alpha}= \Big|\det \Big(\frac{\partial \bar{x}}{\partial x}\Big)\Big|^{-1/4}\frac{\partial \bar{x}^{\mu}}{\partial x^{\alpha}}\label{lormat},
\end{equation}%
which clearly verifies $\Lambda^{\mu}_{\,\alpha}\Lambda^{\alpha}_{\,\nu}=\delta^{\mu}_{\nu}$. Moreover, is easy to see that $\Lambda^{\mu}_{\alpha}$ constitutes an element of the Lorentz group, because, replacing (\ref{lormat}) in (\ref{concaminv}), the conformal factors cancel out and
\begin{equation}
\bar{g}_{\mu\nu} \Lambda^{\mu}_{\,\alpha}\Lambda^{\mu}_{\,\beta}=g_{\alpha\beta}. \label{invde Lor}
\end{equation}%
Eqs. (\ref{det}) and (\ref{lormat}) enable us to tend the bridge between the Lorentz transformations and the special conformal transformations. If a given field $\Psi(x)$ has a well defined behavior under the Lorentz group, we formally have $\Psi'(x')=L(\Lambda)\Psi(x)$, being $L(\Lambda)$ the specific function of the Lorentz transformation represented by $\Lambda\equiv\Lambda^{\mu}_{\,\alpha}=\partial x'^{\mu}/\partial x^{\alpha}$. It follows (see, e.g. \cite{Isham}) that $\Psi(x)$ will transforms under the conformal group according to $\Psi(x)\rightarrow\bar{\Psi}(\bar{x})$, and

\begin{equation}
\bar{\Psi}(\bar{x})=\Big|\det \Big(\frac{\partial \bar{x}}{\partial x}\Big)\Big|^{\ell/4}\,L\Big(\Big|\det \Big(\frac{\partial \bar{x}}{\partial x}\Big)\Big|^{-1/4}\frac{\partial \bar{x}^{\mu}}{\partial x^{\alpha}}\Big)\,\Psi(x), \label{confield}
\end{equation}%
where $\ell$ is a Lorentz scalar called the \emph{weight}. For example, a tensor density of type $B^{\mu\nu}$ and weight $\ell$, will transforms as

\begin{equation}
\bar{B}^{\alpha\beta}(\bar{x})=\Big|\det \Big(\frac{\partial \bar{x}}{\partial x}\Big)\Big|^{(\ell-1)/4}\,\frac{\partial \bar{x}^{\alpha}}{\partial x^{\mu}}\frac{\partial \bar{x}^{\beta}}{\partial x^{\nu}}\,B^{\mu\nu}, \label{transfweightl}
\end{equation}%
which, by (\ref{det}), turns out to be

\begin{equation}
\bar{B}^{\alpha\beta}(\bar{x})=\lambda^{1-\ell}\,\frac{\partial \bar{x}^{\alpha}}{\partial x^{\mu}}\frac{\partial \bar{x}^{\beta}}{\partial x^{\nu}}\,B^{\mu\nu}=\Omega^{(\ell-1)/2}\,\frac{\partial \bar{x}^{\alpha}}{\partial x^{\mu}}\frac{\partial \bar{x}^{\beta}}{\partial x^{\nu}}\,B^{\mu\nu}, \label{transfweightl}
\end{equation}%
where we have used the relation between $\Omega$ and $\lambda$ given in Eq. (\ref{twofunc}). We see, for instance, that the contravariant components of the metric tensor $g^{\mu\nu}$ transform as a tensor density of weight $\ell=-1$ (see Eq. (\ref{concambiocontra})).

With the machinery just developed, we proceed now to prove the conformal invariance of Maxwell's electrodynamics. First, recall that the classical action for the electromagnetic field

\begin{equation}
 I=\frac{1}{2}\int\sqrt{g}\,\Big(\textbf{E}^{2}-\textbf{B}^{2}-2\,\rho\, \phi+2\,\textbf{J}\cdot\textbf{A}\Big)\,d^{4}x\label{lagranmax},
\end{equation}%
can be written in terms of the field strength $F_{\mu\nu}=\partial_{\mu}A_{\nu}-\partial_{\nu}A_{\mu}$ as

\begin{equation}
 I=-\frac{1}{4}\int\sqrt{g}\,\Big(F_{\mu\nu}F^{\mu\nu}+4\,j_{\mu}A^{\mu}\Big)\,d^{4}x\label{lagranmaxcov},
\end{equation}%
where $\sqrt{g}\equiv\sqrt{\mid det(g_{\mu\nu})\mid}$, $A^{\mu}(x)=(\phi,\textbf{A})$ is the electromagnetic four-vector potential, and $j^{\mu}=(\rho,\textbf{j})$ is the charge-current four vector. From the very definition of $F_{\mu\nu}$ we easily obtain

\begin{equation}
\textbf{E}=-\nabla \phi-\partial\textbf{A}/\partial t,\,\,\,\,\,\textbf{B}=\nabla \times \textbf{A}, \label{potenciales}
\end{equation}%
which automatically solve the two homogeneous Maxwell's equations
\begin{equation}
\nabla \times\textbf{E}+\partial\textbf{B}/\partial t=0,\,\,\,\,\,\nabla\cdot\textbf{B}=0. \label{maxhomo}
\end{equation}%
The inhomogeneous equations, which are obtained from the action (\ref{lagranmaxcov}) by varying with respect to the components of the four-vector potential $A^{\mu}(x)$, read
\begin{equation}
\partial^{\mu}F_{\mu\nu}=j_{\nu}, \label{echino}
\end{equation}%
which in the standard notation are
\begin{equation}
\nabla\times\textbf{B}-\partial\textbf{E}/\partial t=\textbf{j},\,\,\,\,\,\nabla\cdot\textbf{E}=\rho. \label{maxinhomo}
\end{equation}%
According to the definition of the covariant derivative necessary to introduce the electromagnetic interactions in the context of the standard model of particle physics, we have $D^{\mu}\equiv\partial^{\mu}-ieA^{\mu}$, an then, $A^{\mu}$ transforms exactly in the same manner as $\partial^{\mu}$\footnote{Of course, the electron charge $e$ is insensible to conformal changes of any sort.}. This means that
\begin{equation}
\bar{A}^{\mu}=\Omega^{-1}(x) \frac{\partial \bar{x}^{\mu}}{\partial x^{\rho}}A^{\rho},\,\,\,\,\,\,\,\bar{A}_{\mu}=\frac{\partial x^{\rho}}{\partial \bar{x}^{\mu}}A_{\rho}, \label{trancuadri}
\end{equation}%
hence, $A^{\mu}$ transforms as a contravariant vector density of weight $\ell=-1$.

The conformal invariance of Maxwell's theory in vacuum is established noting first that $\sqrt{\bar{g}}\,d^{4}\bar{x}=\Omega^{2}\sqrt{g}\,d^{4}x$ (see Eq. (\ref{mapconf})). As $A_{\mu}$ and $\partial_{\mu}$ are conformaly invariant (see Eqs. (\ref{transfveccor}) and (\ref{trancuadri})), then $F_{\mu\nu}$ also is. On the contrary, by virtue of (\ref{concambiocontra}), we have

\begin{equation}
\bar{F}^{\mu\nu}=\bar{g}^{\mu\lambda}\bar{g}^{\nu\sigma}\bar{F}_{\lambda\sigma}=\Omega^{-2}F^{\mu\nu}. \label{transefe}
\end{equation}%
In this way, the conformal factors cancel out in the first term of (\ref{lagranmaxcov}). In order to demand the conformal invariance of the second term in (\ref{lagranmaxcov}), we have to ask that $\bar{j}_{\mu}\bar{A}^{\mu}=\Omega^{-2}\,j_{\mu}A^{\mu}$, which, in turn, leads us to ask
\begin{equation}
\bar{j}_{\mu}=\Omega^{-1} \frac{\partial x^{\rho}}{\partial \bar{x}^{\mu}} j_{\rho},\label{transefe}
\end{equation}%
in view of the transformation law of $A^{\mu}$ in (\ref{trancuadri}). In summary, provided $j_{\mu}$ and $A^{\mu}$ transform in the way they do, the action (\ref{lagranmaxcov}), and then Maxwell's equations (\ref{maxhomo}) and (\ref{maxinhomo}) are insensitive to conformal changes of the metric of the sort (\ref{mapconf}).

\section{Complementary structures and frequency-independent antennas}\label{secciontres}

In this short section we shall briefly discuss the fundamental aspects of self-complementary antennas, and to comment on some very well established results in the field which are essential in the developments of the next section. For a full account see, for instance, \cite{Mushiake2}.

Complementary 2-dimensional structures are defined in the following manner: an infinite plane conducting screen is pierced with apertures of any shape or size, and the resulting screen is called $\mathcal{A}$. Consider then the screen which is obtained by interchanging the region of metal and aperture space in $\mathcal{A}$, and call this second screen $\mathcal{\tilde{A}}$. Then, screens $\mathcal{A}$ and $\mathcal{\tilde{A}}$ are said to be complementary, because, added together they result in a complete, infinite metal screen. Note that we have $\mathcal{A}\cup\mathcal{\tilde{A}}=\mathbb{R}^{2}$, and $\mathcal{A}\cap\mathcal{\tilde{A}}=\emptyset$, so at least one of the structures must be infinite in size. This definition can be extended to 3-dimensional structures as well.

The impedance characteristics of radiating systems constructed upon planar structures and their complementary ones were originally investigated by Booker \cite{Booker}, based on slotted antennas and dipoles. A generalization for complementary antennas of multiple terminals was presented latter in \cite{Deschamps}, and hundreds of works have been published in this field ever since. A particular case of complementary antennas is the self-complementary or mutually dual antennas (proposed long time ago by Mushiake \cite{Mushiake1}), where the structures $\mathcal{A}$ and $\mathcal{\tilde{A}}$ have exactly the same form, and then, both of them are infinite in size. Self-complementary antennas possess the remarkable property of having a constant input impedance $Z_{in}$.

\begin{figure}[ht]
\centering
\includegraphics[scale=.35]{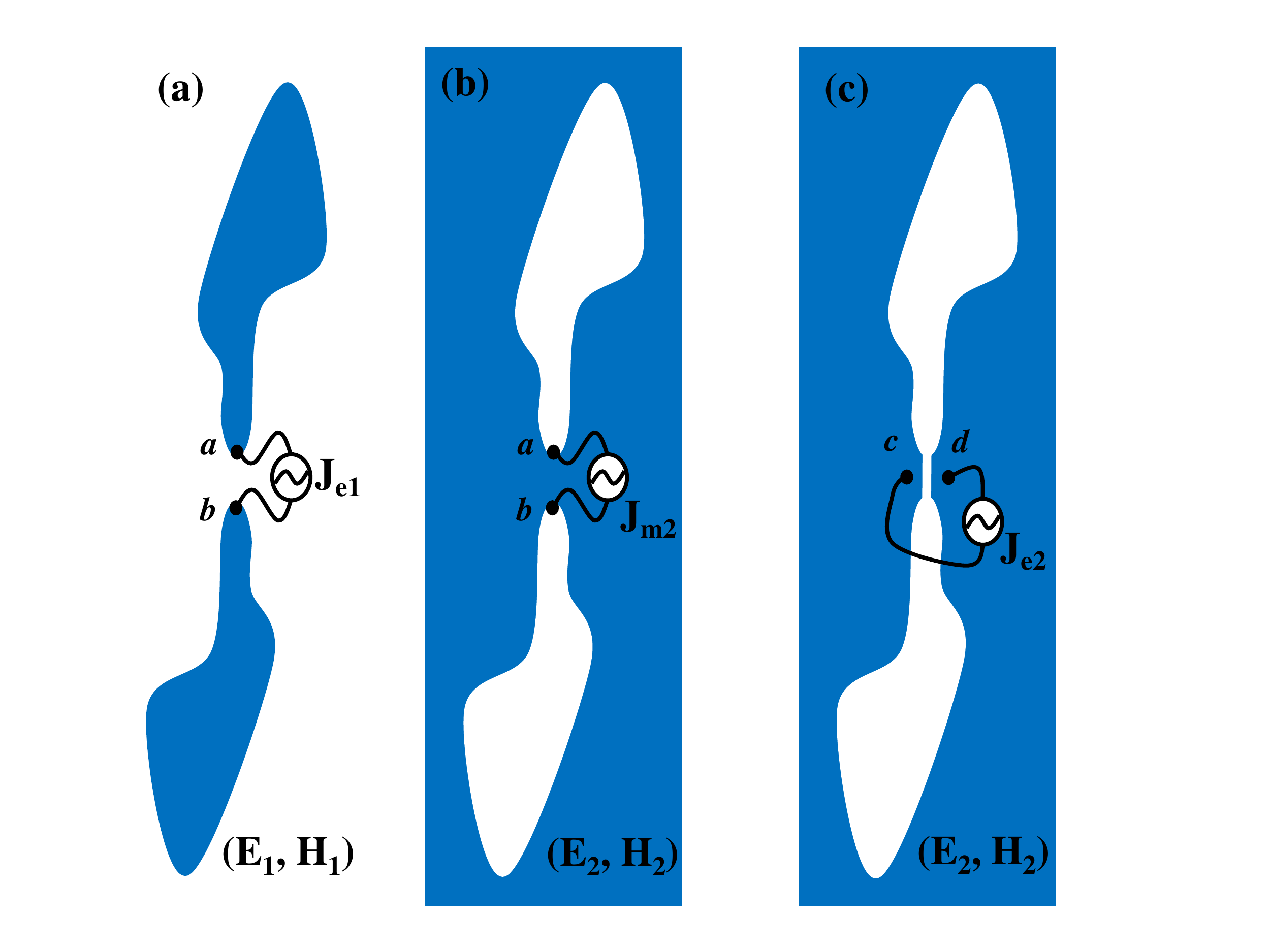} \caption[]{An arbitrary, two arm planar antenna (a), and its complementary (dual) screen, (b) and (c). The screen, constructed out of an infinite plane, can be thought as feeded by means of a magnetic current source (b), or by an electric one as in (c).}\label{estcomp}
\end{figure}

If the electromagnetic fields of an arbitrarily shaped planar antenna and a complementary slot antenna are $(\textbf{E}_{1}, \textbf{H}_{1})$ and $(\textbf{E}_{2}, \textbf{H}_{2})$ respectively (as shown in Fig. (\ref{estcomp})), hence, the complementary character of these two dual structures implies a dual role for the fields involved, namely\footnote{Standard references on this topic are, among many other, Chapter one of \cite{Collin} and Chapter seven of \cite{Balanis}.}

\begin{equation}
\textbf{E}_{2}\leftrightarrow \mp\textbf{H}_{1},\,\,\, \textbf{H}_{2}\leftrightarrow\pm Z^{-2}\, \textbf{E}_{1}, \,\,\,Z^{-2}=\frac{i \omega \epsilon +\sigma}{i \omega \mu},   \label{reldual}
\end{equation}%
where the upper signs are for the front side of the conducting plane, and the lower signs are for its reverse side. In (\ref{reldual}), $Z$ is the impedance of the medium with conductivity $\sigma$, permittivity $\epsilon$ and permeability $\mu$ ($\omega$ is the angular frequency of the fields). Most of the time the complementary slots might be just holes of arbitrary shape in free space, so we have $Z=Z_{0}=(\mu_{0}/\epsilon_{0})^{1/2}$ in that case.

The terminal voltages of these antennas are given by the line integrals of the electric fields in the vicinity of the feeding points, and their input currents are given by the contour integrals around the feeding lines. Therefore, the input impedances of these two antennas, $Z_{in\,1}$ and $Z_{in\,2}$, can be expressed by the ratios of those integrals. Strictly, for the input impedance to have meaning, it is necessary to consider an infinitesimal gap at the feeding point. The voltage at the terminals is given by
\begin{equation}
V_{in}=-\int_{a}^{b} \textbf{E}\cdot\textbf{dl},   \label{voltin}
\end{equation}%
which has an unambiguous significance in view that $a\approx b$. The input current $I_{in}$ can be obtained from the Ampere-Maxwell law (first equation in (\ref{maxinhomo}))
\begin{equation}
I_{in}=\int_{S}\textbf{J}_{in}\cdot\textbf{dS}=\oint\textbf{H}\cdot\textbf{dl}=2\int_{c}^{d} \textbf{H}\cdot\textbf{dl}.  \label{voltin}
\end{equation}%
It is important to mention that the contribution coming from the Maxwell's displacement current $\textbf{J}_{d}=\partial\textbf{E}/\partial t$ is null, because it involves a flux integral $\int\textbf{J}_{d}\cdot\textbf{dS}$ that vanishes by virtue of the infinitesimal character of the gap at the feeding point. The input impedances of both structures are then

\begin{equation}
Z_{in\,1}=\frac{\int_{a}^{b}\textbf{E}_{1}\cdot\textbf{dl}}{2\int_{c}^{d} \textbf{H}_{1}\cdot\textbf{dl}}, \,\,\,Z_{in\,2}= \frac{\int_{c}^{d}\textbf{E}_{2}\cdot\textbf{dl}}{2\int_{a}^{b} \textbf{H}_{2}\cdot\textbf{dl}}.\label{zinboth}
\end{equation}%
By taking the product of the expressions for these two impedances and introducing the relations given in (\ref{reldual}), it is very simple to show that $Z_{in\,1}Z_{in\,2}=Z^{2}/4$. In the particular case of self-complementary antennas $\mathcal{A}=\mathcal{\tilde{A}}$, and then the input impedance obeys \emph{Mushiake's relation}
\begin{equation}
Z_{in}=\frac{1}{2}\Big(\frac{i \omega \mu }{i \omega \epsilon +\sigma}\Big)^{1/2},\label{Mushiake}
\end{equation}%
which, in free space, reduces to the simple expression
\begin{equation}
Z_{in}=\frac{1}{2}Z_{0}=60 \pi\, [Ohm].\label{Mushiakefree}
\end{equation}%
These last two relations mean that the input impedance of a self complementary antenna is constant, irrespective of its specific shape and operation frequency.

\section{The novelty: conformal edge antennas}\label{seccioncuatro}

The current distributions on planar infinite sheets out of which a given self-complementary antenna is theoretically conceived, are disturbed by the truncation which is necessarily introduced in order to implement it in practice. For instance, there will be some harmful effects at the feeding point as a result of reflected currents from the truncated ends. Such truncation effects necessarily cause deterioration of the constant-impedance property assured by the theory.
In this section, we propose an specific family of conformal factors $\Omega(x)$ as emerge from the analysis of the conformal structure of infinity performed mainly by Penrose in Refs. \cite{Penrose1} and \cite{Penrose2}. \emph{Conformal edge antennas}\footnote{Here \emph{conformal} refers to the group briefly exposed in Sec. \ref{secciondos}, and not to  ``adapted to a given surface'', as in Ref. \cite{ConformalAnt}.} will then be obtained from usual planar, self-complementary structures by applying these kind of transformations to the points lying on the arms of the latter.

In particular, bearing the application we have in mind, it is sufficient to deal with the conformal structure of flat Minkowski spacetime in four dimensions. The novel idea we propose, is to combine these tools together with the conformal invariance of $D=4$ electrodynamics to obtain compact (finite) antennas which, in certain specific sense, capture the properties of spatial infinity by means of an intelligent (and physically motivated), choice of the conformal factor $\Omega(x)$.

In order to further pursuit this goal we will give a brief account of the procedure of attaching to Minkowski space a collection of special points representing time and spatial infinity, as developed in the above mentioned references, where we refer the reader for a thorough exposition. To see how this works, consider the line element of Minkowski spacetime in pseudo-euclidian coordinates $(t,x,y,z)$,  $ds^{2}=dt^{2}-dx^{2}-dy^{2}-dz^{2}$, but written in \emph{advanced} and \emph{retarded} null coordinates $v=t+r$, $u=t-r$ (note that $u\leq v$), where $r^{2}=x^{2}+y^{2}+z^{2}$ and, as usual,
\begin{equation}
x=r \,cos\theta\, cos\phi,\,\,\,\,y=r\, sin\theta\, cos\phi,\,\,\,\,z=r\, sin\phi.  \label{coordesfe}
\end{equation}%
In these coordinates $(u,v,\theta,\phi)$, the metric adopts the form
\begin{equation}
ds^{2}=dudv-dr^{2}-\frac{1}{4}(u-v)^{2}(d\theta^{2}+sin^{2}\theta\, d\phi^{2}).\label{minknulas}
\end{equation}%
A further coordinate change will bring up the conformal structure explicitly. Let be $(p,q,\theta,\phi)$ the coordinates defined by
\begin{equation}
v=tan\,p,\,\,\,\,u=tan\,q, \label{coorconf}
\end{equation}%
so we have $-\pi/2\leq q\leq p\leq\pi/2$. This change has the effect that the points at infinity have finite values as viewed in $p,q$ coordinates. Using that $dv=sec^{2}p\,dp$ and $du=sec^{2}q\,dq$, we have that (\ref{minknulas}) results

 \begin{equation}
ds^{2}=sec^{2}\,p\,sec^{2}\,q\,d\bar{s}^{2},\label{intconf}
\end{equation}%
where $d\bar{s}^{2}=dpdq-\frac{1}{4}sin^{2}(p-q)(d\theta^{2}+sin^{2}\theta\, d\phi^{2})$, or, in the notation we have adopted from the outset (see Eqs. (\ref{difcuadra}) and (\ref{twofunc})),

\begin{equation}
d\bar{s}^{2}= \Omega(p,q)\,ds^{2},\,\,\,\,\,\Omega=\lambda^{-2}=sec^{-2}\,p\,sec^{-2}\,q. \label{final}
\end{equation}%
The important point is that the metric $d\bar{s}^{2}$ is perfectly regular at $p=\pi/2$ and $q=-\pi/2$ (whereas $ds^{2}$ is not, because those points correspond to $v=\infty$ and $u=-\infty$ in that space-time). The situation is such that we have a well-defined conformal structure on a manifold (given by $-\pi/2\leq q\leq p\leq \pi/2$) with boundary (given by $q=-\pi/2$ or $p=\pi/2$), and its interior (given by $-\pi/2<q\leq p<\pi/2$) is identical in conformal structure with Minkowski space-time. This means that we have attached to Minkowski space-time a boundary which consists of points which are actually at infinity, obtaining in this way a new manifold which is compact\footnote{This procedure is, however, non unique. In fact, the kind of compact manifold obtained comes from the specific coordinate change (\ref{coorconf}). Other changes are admissible in order to get a boundary, for instance $v=tan(p+p_{0}\,p^{n}), u=tan(q+q_{0}\,q^{n})$, with $p_{0},q_{0}\geq0$ and any odd natural number $n$. What makes (\ref{coorconf}) so important, is that the $d\bar{s}^{2}$ below Eq. (\ref{intconf}) obtained by these means \emph{also} constitutes a solution of the equations of motion of Einstein's General Relativity, called \emph{Einstein's static universe}. Even though all this is clearly beyond the scope of the present article, the interested reader is invited to consult Chapter five of \cite{H-E} for further details.}.

The standard truncation procedure used in practice can be summarized as follows:

\begin{description}
  \item[$(\ast)$] Take a 2D self-complementary antenna $\mathcal{A}$ of an arbitrary shape and infinite dimensions, and set a planar circular region of the space of radius $R$ centered at the feeding point of $\mathcal{A}$. This circular region will be a measure of the size of the antenna, possibly imposed by a given application. Then, cut $\mathcal{A}$ in such a way that it ends up being totally circumscribed in the disk of radius $R$, obtaining thus a new truncated, physically realistic antenna $\mathcal{A}_{T}$.
\end{description}
The main idea we propose in order to implement a more clever truncation can be stated, in turn:
\begin{description}
  \item[$(\ast\ast)$] Take into account the infinite region previously left aside in the truncation process of $\mathcal{A}$, and perform a special conformal transformation to it in order to bring infinity up to the compact region whose boundary is the circle of radius $R$. This can be done by choosing the conformal factor $\Omega(x)$ in an appropriated way which, in turn, might be highly non unique. The effect of the conformal transformation is to squash everything up near infinity, thus, points close to the circle are actually very distant in the usual (untransformed) picture. The antennas $\mathcal{A}_{\Omega}$ so obtained will have all better performances concerning the constancy of $Z_{in}$, even though this will depend on the choices of $\Omega$.
\end{description}

The property that assures a better behavior of $Z_{in}$ for $\mathcal{A}_{\Omega}$ is just the invariance of the former under conformal changes of the metric. We proceed now to expose and discuss this matter.

As mentioned before, the input impedance $Z_{in}$ can be expressed by the ratio of the input voltage to the input current at the feeding point.
Of course, expressions of the sort (\ref{zinboth}) do not take into account the finiteness of the actual antenna, and so, they constitute just an approximation in a real truncated system. Nonetheless, we can make use of a conformal change in order to bring infinity up to a finite distance, and then, to perform a more intelligent truncation with the aim to capture some of the properties left behind in the cutting. Clearly, for this process to be meaningful, we have to show first that $Z_{in}$ is conformaly invariant.

In general, when the scale factor is an arbitrary smooth and non null function, the invariance of $Z_{in}$ must be established on the basis of the transformation law for $\textbf{E}$ and $\textbf{H}$, as well as the differential $\textbf{dl}$ involved in expression (\ref{zinboth}). In turn, and for the sake of simplicity, we can elaborate an heuristic argument in order to prove the covariance of (\ref{zinboth}). As mentioned in the preceding section, the integrals in (\ref{zinboth}) are performed along paths surrounding the feeding terminals, which are very close to each other. In other words, the integrals should be calculated along paths which can be embedded in a very small sphere of radius $r_{0}$ centered at the feeding  point. In view that the conformal change just alters the structure of the space in the remote regions away from the origin, we can simply think about it as if the conformal factor behaves in a neighborhood of the origin as $\Omega(r)=r+\mathcal{O}(r^{2})$. This means that in a small neighborhood of the terminals, every field is invariant under the conformal change, and then, so it is $Z_{in}$ in equation (\ref{zinboth}). Although that these considerations repose in a choice of a specific family of conformal factors (all of them smoothly approaching the identity at the origin), they are perfectly pertinent to the present purposes.

\section{Engineering matters and results}\label{seccioncinco}

We proceed now to implement the conformal techniques exposed above in a couple of designs which are of extensive use nowadays in UWB communications systems. The idea is to modify the structure of the space-time --and then, everything within it-- with the aim of distorting the usual notion of distance between points as we know from our contact with the euclidian world. As explained in \ref{seccioncuatro} this will allows us to give precise meaning to sentences of the kind "to cut a hole out from an infinite conducting plane" or "at an infinite distance away from the sources", which are in the sine of many of the idealizations behind electrodynamics.

We shall concentrate our efforts on one specific self complementary, two-dimensional planar radiating systems. Specifically, we shall deal as an example with the logarithmic spiral antenna. Due to the fact that our radiating structures are not changing in time, we will slice the space-time in constant time hypersurfaces, and to place our antennas at $t=0$ without loss of generality. Besides, in view that we are dealing with planar structures, we shall set the polar angle $\phi$ of an spherical coordinate system $(r,\theta,\phi)$, to the value $\phi=\pi/2$. So, effectively, we will describe our antennas in terms of polar coordinates $(r,\theta)$.

Let $\mathcal{A}_{1}$ and $\mathcal{A}_{2}$ be the regions in $\mathbb{R}^{2}$ delimited by the polar curves

\begin{equation}
\left\{
  \begin{array}{ll}
    \mathcal{A}_{1}  \\
    \mathcal{A}_{2}
  \end{array}
\right.=\left\{
           \begin{array}{ll}
             \langle r_{1}=r_{0}\, Exp\Big(\alpha\,\theta+\langle^{\pi/2}_{0}\Big) \\
             \langle r_{2}=r_{0}\, Exp\Big(\alpha\,\theta+\langle^{3\pi/2}_{\pi}\Big)
          \end{array}
         \right.,
 \label{espilog}
\end{equation}%
where $r_{0}$ and $\alpha$ are two constants representing the initial distance to the origin and the growing rate of the spiral, respectively. In (\ref{espilog}), (let us say) the symbol $\langle r_{1}$ actually represents two curves, one with a null phase, and the other with a phase of $\pi/2$. The two-arms self-complementary, planar logarithmic spiral is defined according to $\mathcal{A}=\mathcal{C}\ell(\mathcal{A}_{1}\cup\mathcal{A}_{2})$ (here $\mathcal{C}\ell$ refers to the mathematical closure of a set). Note that $\mathcal{A}_{2}$ is just $Rot_{\pi}\,\mathcal{A}_{1}$ (a rotation of $\pi$ radians about an axis orthogonal to the plane of the spiral), and $\mathbb{R}^{2}-\mathcal{A}=\,Rot_{\pi/2}\,\mathcal{A}$. In this way, the logarithmic spiral just defined is actually self-complementary.

\bigskip

\begin{figure}[ht]
\centering
\includegraphics[scale=.26]{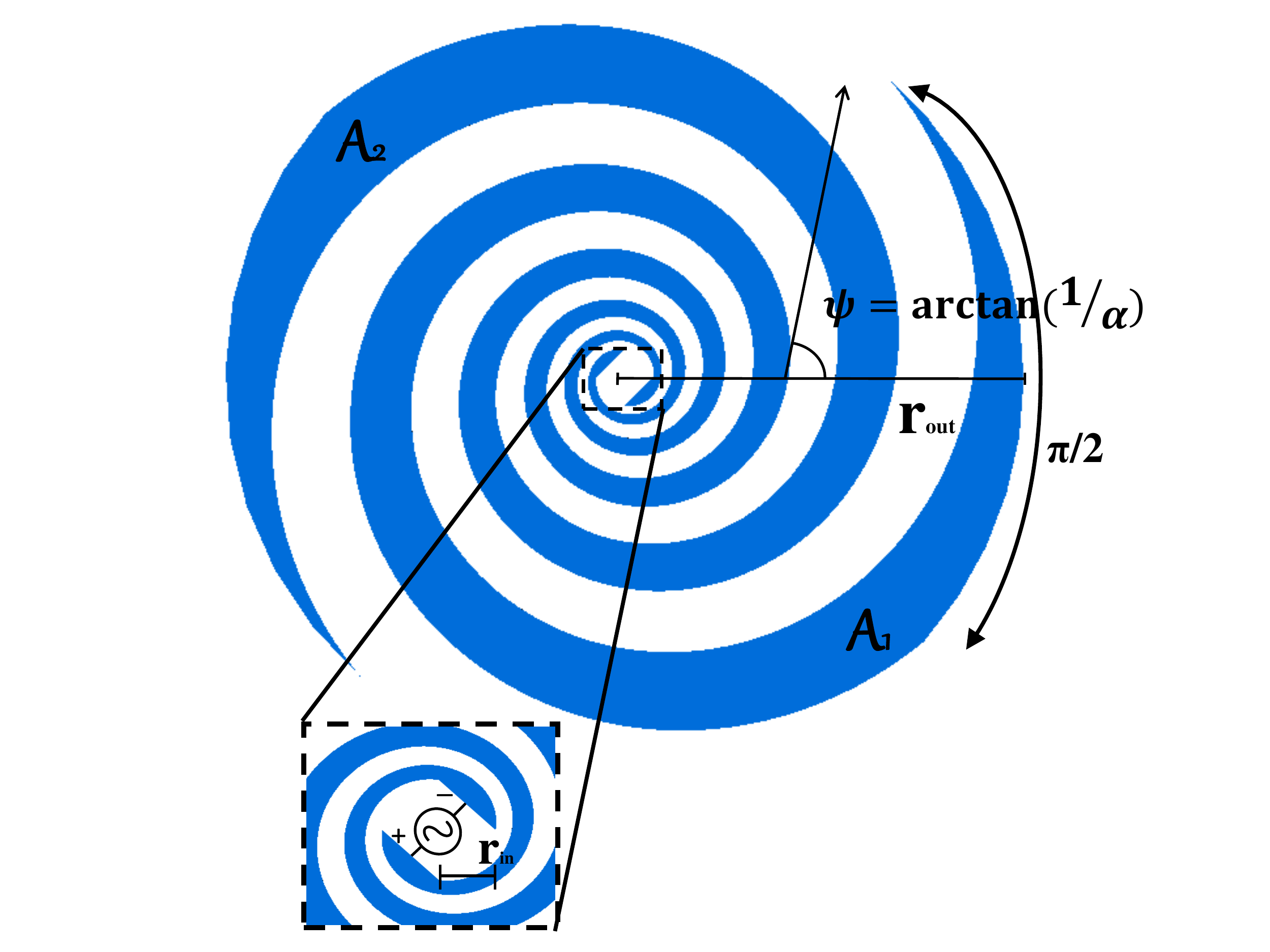}  \caption[]{The logarithmic spiral antenna, as defined in eq. (\ref{espilog}), and subjected to the standard truncation procedure.}\label{figespi}
\end{figure}

The exposed design, as well as its intervening parameters, is depicted in Fig. \ref{figespi}. Clearly, the model appearing in Fig. \ref{figespi} was affected by a truncation process of the kind $(\ast)$, as explained in the previous section. Our aim now is to conceive some conformal transformations in order to obtain a family $\mathcal{A}_{\Omega}$ of conformal edge antennas of the sort envisioned in $(\ast\ast)$.

Any point $p_{0}$ of the regions $\mathcal{A}_{1}$ and $\mathcal{A}_{2}$ of the antenna can be thought as the image of a given value of the polar angle (let us say, $\theta_{0}$), under a curve $r(\theta)$ lying in the corresponding region. Of course, $r(\theta_{0})$ also represents the distance between $p_{0}$ and the origin, then, a conformal transformation squashing distances will act directly on the radial coordinate in the way $r(\theta)\rightarrow \bar{r}(\theta)=f(r(\theta))$, for some suitable function $f(r)$. If we follow the procedure exposed in \ref{seccioncuatro}, the advanced and retarded coordinates become $v=r$, $u=-r$ (note that $t=0$), so they are not actually independent on the hypersurface $t=0$, which in turn, seals also the dependency of the coordinates $p,q$ defined in Eq. (\ref{coorconf}). As a matter of fact we also have $p=-q$ when $t=0$, and the relevant conformal coordinate change is $p\equiv \bar{r}=\arctan(r)$, so the function $f$ just mentioned results $f(r)=\arctan(r)$ in this particular case, and the resulting conformal edge antenna $\mathcal{A}_{\Omega}$ reads

\begin{equation}
\mathcal{A}_{\Omega}=\mathcal{C}\ell(\mathcal{A}_{\Omega\,1}\cup\mathcal{A}_{\Omega\,2}),\,\,\,\,\,\left\{
  \begin{array}{ll}
    \mathcal{A}_{\Omega\,1}  \\
    \mathcal{A}_{\Omega\,2}
  \end{array}
\right.=\left\{
           \begin{array}{ll}
             k_{0}\arctan(\langle r_{1}) \\
             k_{0}\arctan(\langle r_{2})
          \end{array}
         \right.,\label{confedge}
\end{equation}%
where $\langle r_{1}$ and $\langle r_{2}$ are defined in Eq. (\ref{espilog}), and $k_{0}$ is a scaling factor with units of length. It is clear that the conformal edge antennas obtained by this procedure are compactly supported, in the sense that they possess a maximum external radius $r_{out}$ of $k_{0}\pi/2$. This external radius is representative of the set of points which live at spatial infinity, and the entire circle of radius $r_{out}$ represents those points which in our usual picture are obtained by means of the limit $r\rightarrow\infty$.

\bigskip

\begin{figure}[ht]
\centering
\includegraphics[scale=.33]{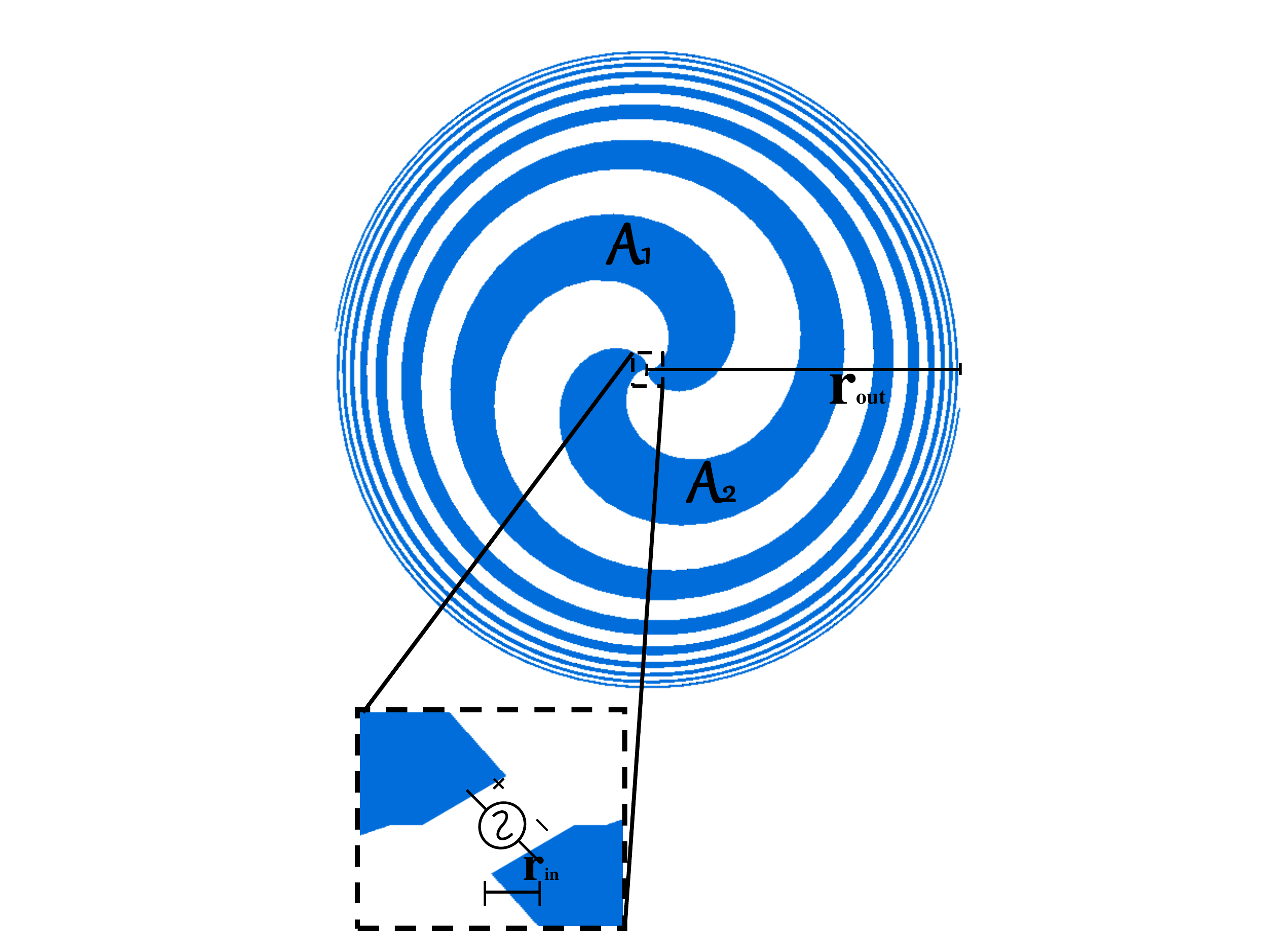}  \caption[]{A conformal change was applied to the standard spiral antenna of eq. (\ref{espilog}), obtaining in this way a type of \emph{conformal edge spiral} antenna.}\label{metodoconforme2}
\end{figure}

The notion of compactness behind conformal edge antennas is not patrimony of the specific coordinate change $\bar{r}=\arctan(r)$. We can try other options with the purpose of putting a bound to the radial coordinate. Among a variety of choices, we can rehearse $\alpha \theta\rightarrow \theta'=\arctan(\alpha \theta)$. In this case we would have

\begin{equation}
\mathcal{A}_{\Omega}=\mathcal{C}\ell(\mathcal{A}_{\Omega\,1}\cup\mathcal{A}_{\Omega\,2}),\,\,\,\,\,\left\{
  \begin{array}{ll}
    \mathcal{A}_{\Omega\,1}  \\
    \mathcal{A}_{\Omega\,2}
  \end{array}
\right.=\left\{
           \begin{array}{ll}
             \langle r_{1}(\theta') \\
             \langle r_{2}(\theta')
          \end{array}
         \right.,\label{confedgebis}
\end{equation}%
The idea behind the transformations of the kind (\ref{confedge}) and (\ref{confedgebis}) is schematically depicted in Fig. \ref{metodoconforme2}. It is important to mention that, despite the resemblance with some previous developments (see \cite{Power1} and \cite{Power2}), the antennas here exposed are based on first principles and in conformal methods which are applicable to any self-complementary radiating element, even in three dimensional space and to antennas of an arbitrary structure and geometry.

\begin{figure}[ht]
\centering
\includegraphics[scale=.18]{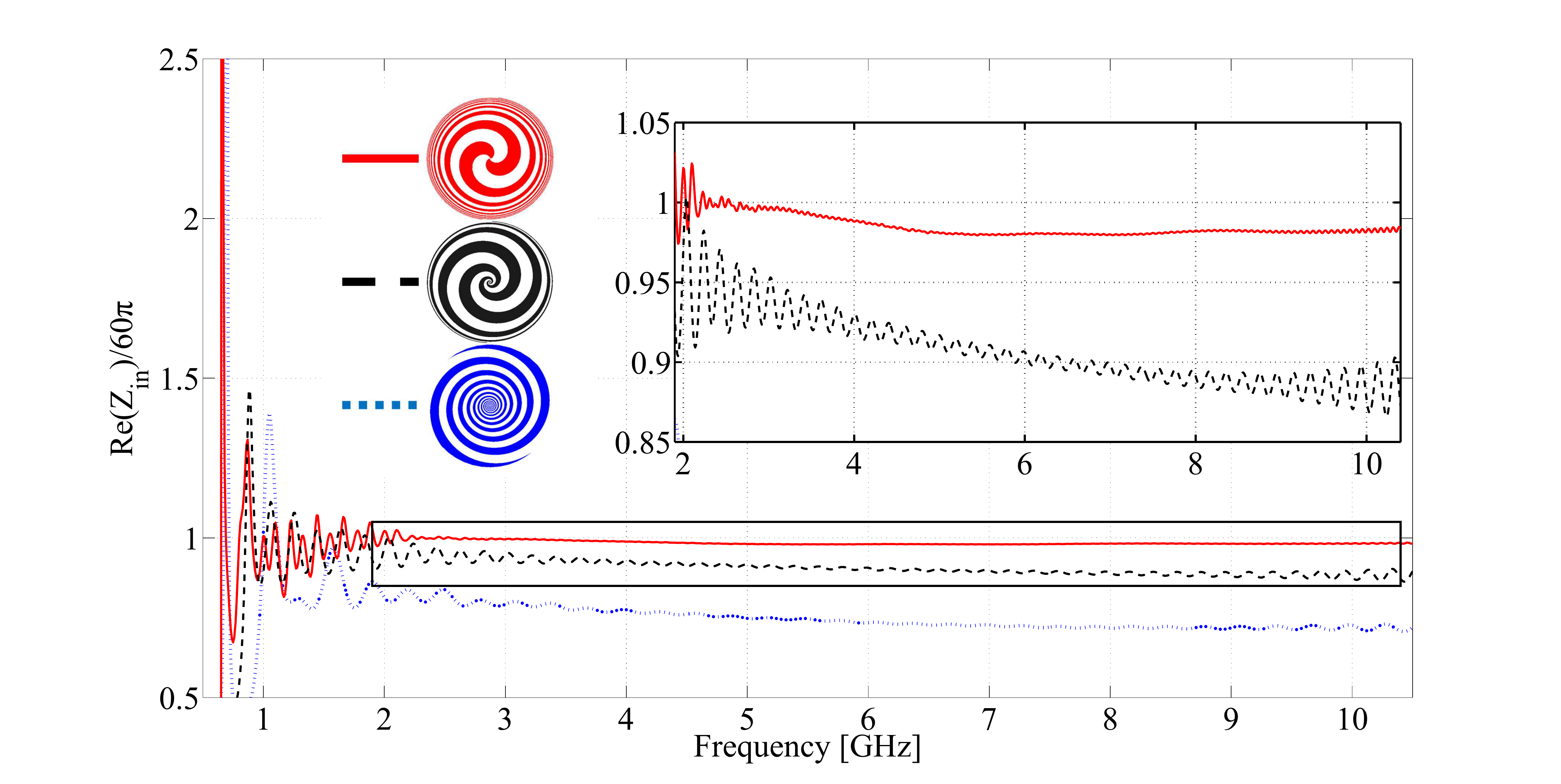} \caption[]{The normalized real part of the input impedance $Re(Z_{in})/60\pi$ as a function of the frequency for three planar antennas. Apart from the standard logarithmic spiral (blue dotted curve), we show two conformal edge spiral antennas, CESP1 (black dashed) and CESP2 (red solid).}\label{comptresz}
\end{figure}

\begin{figure}[ht]
\centering
\includegraphics[scale=.18]{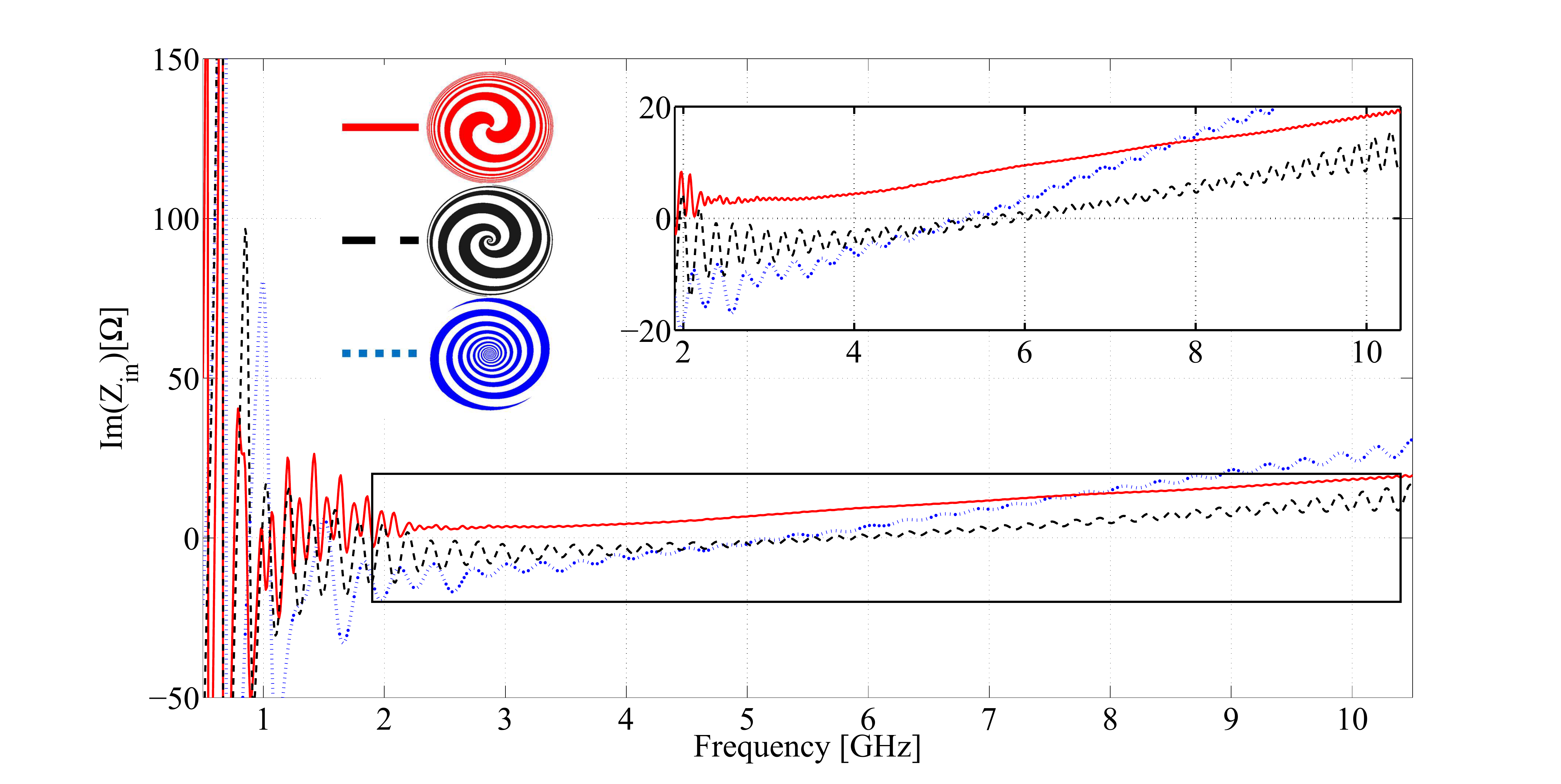} \caption[]{The imaginary part of the input impedance as a function of the frequency for the same designs as in Fig. (\ref{comptresz}).}\label{comptreszima}
\end{figure}

We now proceed to simulate several designs, and to characterize the input impedance of them. In Figs. \ref{comptresz} and \ref{comptreszima} the real and imaginary parts of the input impedance as a function of the frequency for three different planar antennas with the same inner and outer radius of $1 mm$ and $70 mm$ respectively, operating in the S-C-X bands, is shown. To require the same size in the three proposed designs, obviously involves different choices of the intervenient parameters $\alpha$, $k_{0}$ and $r_{0}$.

Compared with a standard logarithmic spiral (blue dotted curve), we have two different conformal edge spirals coming from different conformal changes. The one in dashed black line was obtained from the spiral by means of a conformal change of the form (\ref{confedge}), while the solid red curve was constructed upon a conformal transformation of the sort (\ref{confedgebis}). We shall call these two antennas \emph{conformal edge spirals} (CESP).

Even though all three designs have nice constant input impedance performances in a range of about 8 GHz (from 2 GHz to 10 GHz), we clearly see that the conformal ones approach the theoretical value $Re(Z_{in})=60\pi$ predicted by Mushiake's relation (\ref{Mushiake}) in a higher degree of exactness. As a matter of fact, the constant impedance property is remarkably achieved within less than a 15\% for CESP1 and within less than 3\% for CESP2 along the 8 GHz range mentioned. The frequency-independent behavior of the CESP's is also evident in the more controlled value of the $Im(Z_{in})$, as witnessed in Fig. \ref{comptreszima}, which shows that it varies just up to 20 Ohms in the whole operation range, comparing with a variation of about 50 Ohms of the usual spiral along the same frequency range.

\begin{figure}[ht]
\centering
\includegraphics[scale=.18]{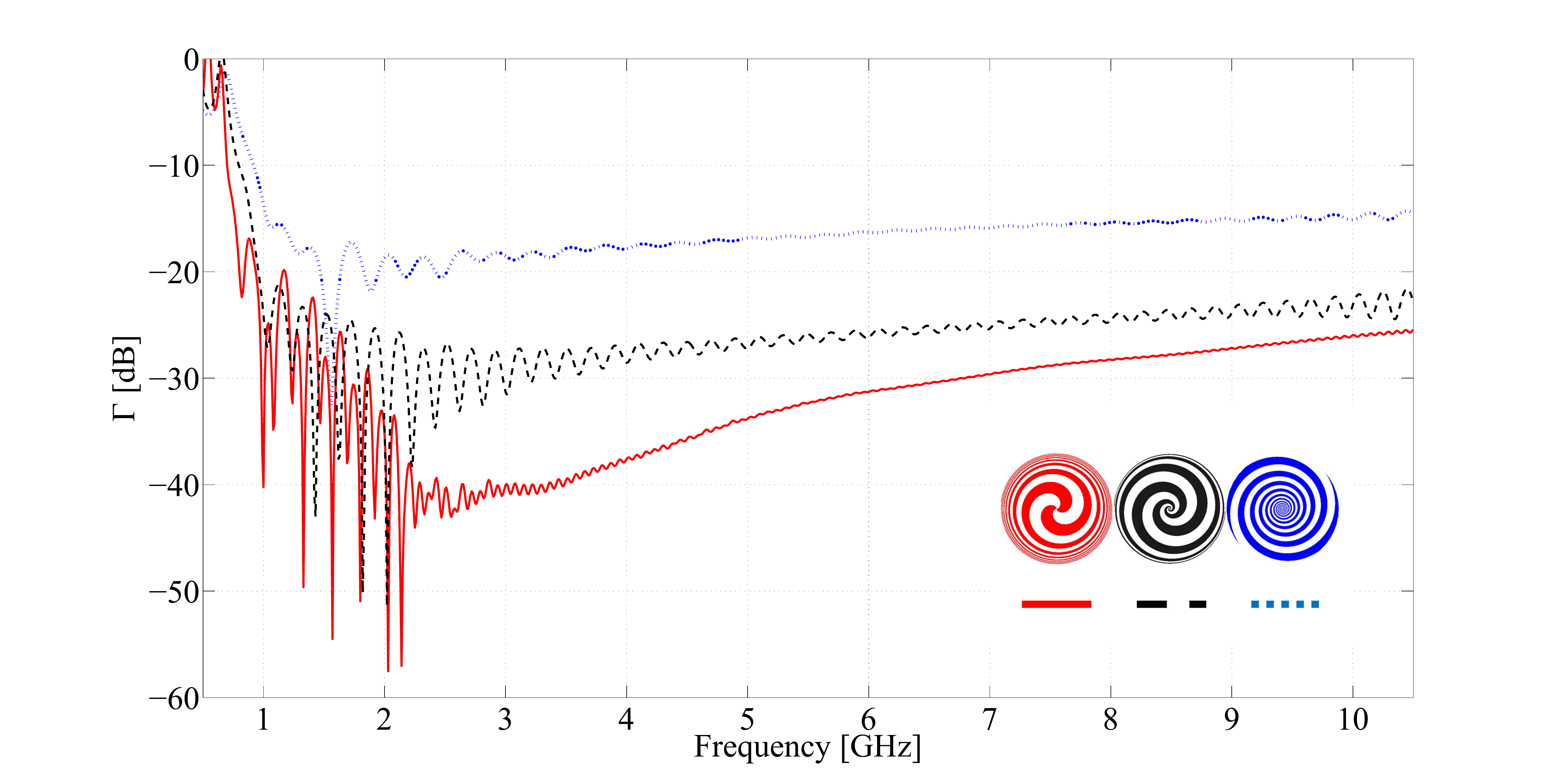} \caption[]{Reflexion coefficient as a function of the frequency for the same designs as in Fig. (\ref{comptresz}).} \label{gamadef}
\end{figure}

An alternative visualization of the fine performances of the CESP's so designed is given by the amount of reflected power as a consequence of the mismatch between the impedance of the feeding port (taken as $60\pi$), and the input impedance $Z_{in}$ of the antennas. The reflection coefficient $\Gamma$ is depicted in Fig. \ref{gamadef}, where the standard $-10$ dB marginal value, achieved by the logarithmic spiral, is considerably improved to less than $-20$ dB in a wider range of more than 9 GHz by the two CESP's. Note that the $-10$ dB value is obtained in the CESP's at frequencies as low as about 750 MHz.

\begin{figure}[ht]
\centering
\includegraphics[scale=.18]{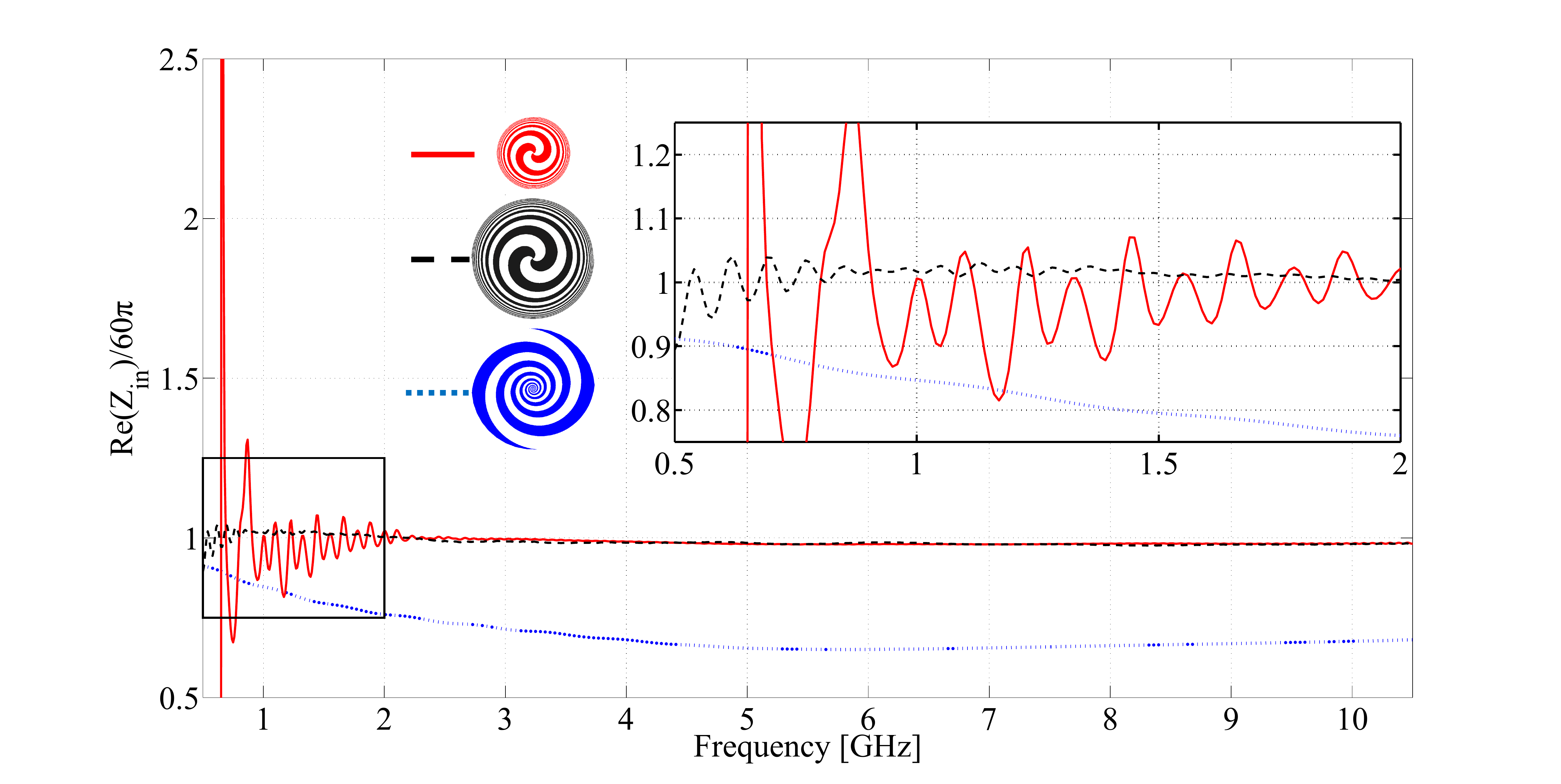} \caption[]{Normalized real part of the input impedance as a function of the frequency for a double radius spiral (dotted blue) and double radius conformal edge spiral (CESP3, dashed black), compared with the CESP2 of Fig. \ref{comptresz}.} \label{compdoblereal}
\end{figure}

\begin{figure}[ht]
\centering
\includegraphics[scale=.18]{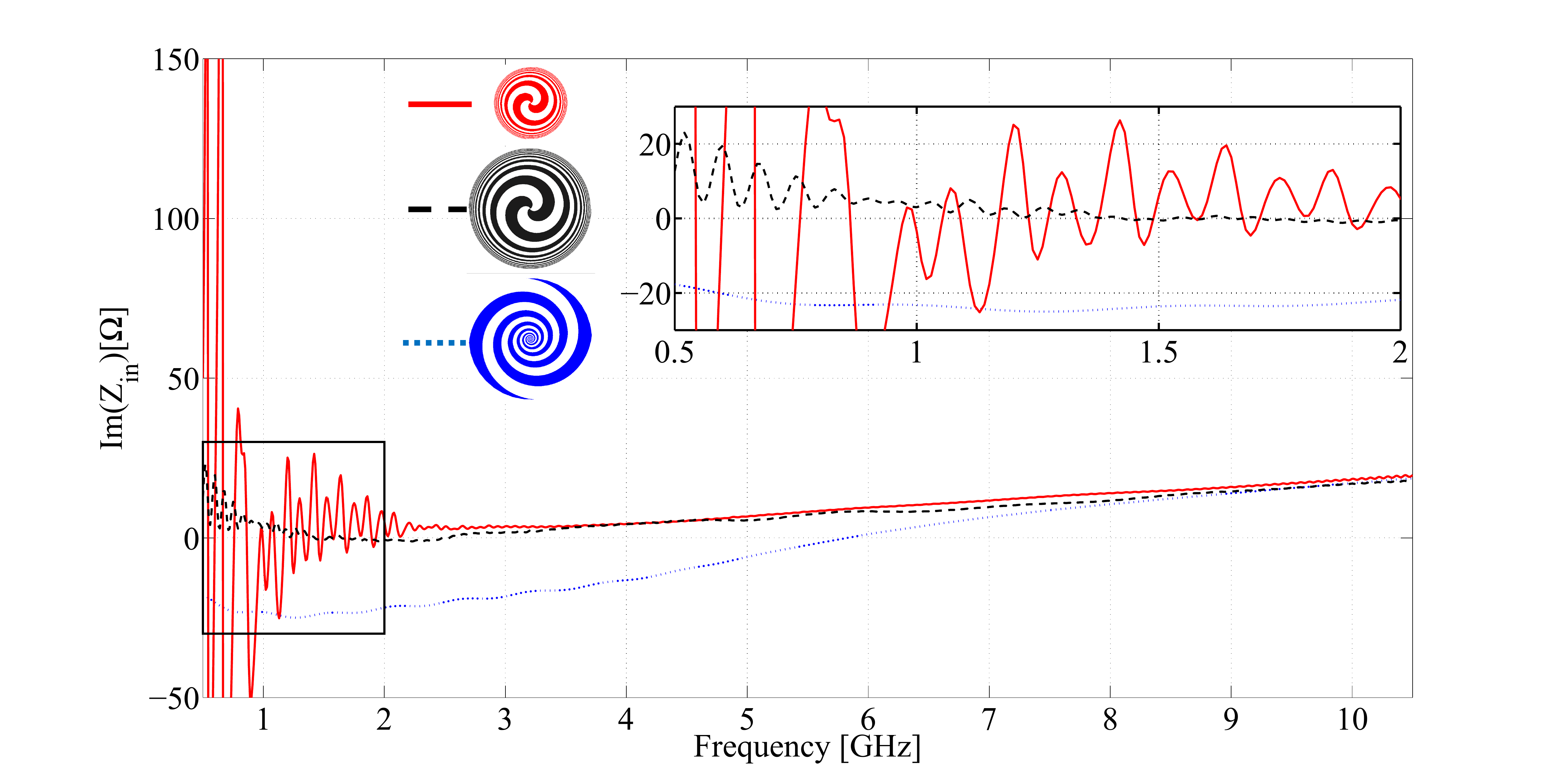} \caption[]{Imaginary part of the input impedance as a function of the frequency in the same circunstances as in the previous figure.}\label{compdobleimag}
\end{figure}

\begin{figure}[ht]
\centering
\includegraphics[scale=.18]{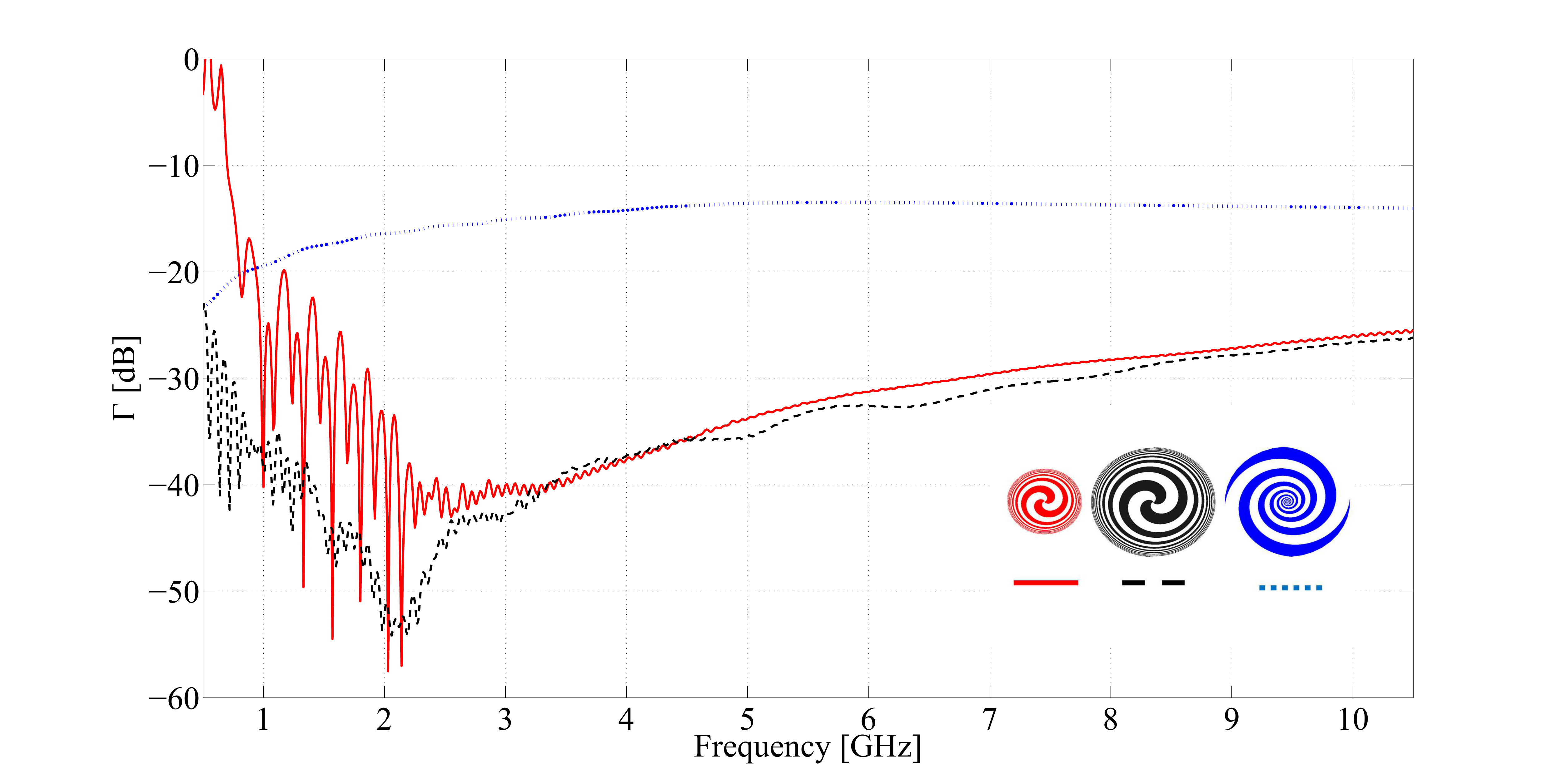} \caption[]{Reflexion coeficcient as a function of the frequency for the same designs as in Fig. \ref{compdoblereal}.}\label{otrogamma}
\end{figure}

If the application requires to further extend the bandwidth towards the low frequency regime, an enlargement of the dimensions of the antennas is necessary. Nonetheless, the technique here developed allows to expand the bandwidth by keeping the radiating system as small as possible. Figs. \ref{compdoblereal} and \ref{compdobleimag} show the real and imaginary part, respectively, of the input impedance of two additional antennas, compared with the CESP2 worked before. Now, we have in dashed black a new conformal edge spiral (CESP3) with an outer radius of $140 mm$ (twice the one of CESP2, maintaining the inner radius of $1mm$ and exactly the same feeding port as before), and a new spiral (in dotted blue) of the same dimensions than those of CESP3. It is clear that even though the size of the spiral doubles the one of CESP2, its performance is better just below 1 GHz, where the impedance of CESP2 has a strongly oscillating character due to the fact that the ratio between the wavelength and the diameter of the antenna is $\lambda/D\approx2$. In total contrast and, despite of having the same outer radius than the logarithmic spiral, the input impedance of CESP3 differs from $60\pi$ in less than $5\%$ in the range 0.6 - 10 GHz, and its performance is very nice even at frequencies as low as 500 MHz, where $Re(Z_{in})/60\pi\approx 0.9$. These fine features are accompanied by the corresponding proper behavior of $Im(Z_{in})$ (see Fig. \ref{compdobleimag}), and also by the reflection coefficient shown in Fig. \ref{otrogamma}.

\section{Conclusions}\label{conclusiones}

In this article we have gathered several techniques coming from quite different branches of physics and mathematics, such as conformal field theory, electrodynamics and differential geometry, and combined them in order to get a consistent framework capable of dealing with certain practical issues in the field of UWB, self-complementary planar antennas. After reviewing the fundamentals of the conformal group and its relation with classical electrodynamics, we have briefly introduced frequency-independent antennas and their constant input impedance property as it comes from Mushiake's relation, Eq. (\ref{Mushiake}). This theoretical relation is strongly based on the fact that (planar) SC antennas are constructed out of an infinite conducting plane. However, actual SC antennas are always subjected to a truncation process of some kind which crucially affects their performances in general, and their input impedances in particular. Bearing this unavoidable limitation in mind, we have asked under what circumstances a different truncation procedure can be implemented on a given SC antenna in order to have a more controlled input impedance level along the widest possible frequency range.

In section \ref{seccioncuatro} we gave a definite answer to this question. Applying --as Penrose did-- certain conformal transformations to the flat (Minkowskian) space-time, we have shown how a given antenna can be designed to have a radius which is just its \emph{conformal edge}. This boundary is, theoretically,  representative of spatial infinity, so we have in practice a compact antenna which actually behaves as if it were bigger. Regarding this, section \ref{seccioncinco} was devoted to the design and simulation of one specific example of conformal edge antenna, namely, the one coming from a conformal transformation applied to a SC logarithmic antenna, which we have called conformal edge spiral (CESP). We have opportunely seen that the improvements predicted by the theory concerning the constancy of the input impedance curve, are confirmed by the simulations. In particular, depending on the size and shape of the CESP's, some values of $Re(Z_{in})$ differ from $60\pi$ within less than 5 $\%$ along a frequency range of 0.6 - 10 GHz (CESP3, see Figs. \ref{compdoblereal} and \ref{compdobleimag}). Particularly remarkable, in view of its small dimensions, is the performance of CESP2 (see. Figs. \ref{comptresz} and \ref{comptreszima}), whose input impedance $Re(Z_{in})/60\pi$ ranges between $0.97$ and 1, along 8 GHz in the band $2-10$ GHz. It will be matter of future works to construct and characterize several prototypes of this kind.

The framework here exposed has a number of natural applications that we have not discussed, which go far beyond planar SC radiating systems, and that will constitute material for future developments. Among them we can mention the generalization to SC structures in flat three spatial dimensions, the analysis of the input impedance of antennas constructed on curved surfaces of arbitrary shape, and the study of the conformal structure of arrays of radiating systems. On more formal grounds, it would be also interesting to approach the study concerning the limitations of a given radiating system in general --along the lines of those discussed in \cite{Gustav}-- in the light of the ideas here exposed.

\bigskip

\emph{Acknowledgments} This work was supported by Consejo Nacional de Investigaciones Científicas y Técnicas (CONICET), Instituto Balseiro and Comisión Nacional de Energía Atómica (CNEA). F.F, and P.C.C are members of Carrera del Investigador Científico (CONICET).


\begin{thebibliography}{99}

\bibitem{UWB1} Reed J, 2005, \emph{Introduction to Ultra Wide Band Communication Systems}. Prentice-Hall, Englewood Cliffs, NJ.

\bibitem{UWB1b} Bond E J, Li X, Hagness S C, and Van Veen B D, 2008, \emph{Microwave imaging via space-time beamforming for early detection of breast cancer}. IEEE Trans. Antennas Propag. \textbf{51}, 1690.

\bibitem{UWB1c} Chen M, Gonzalez S, Vasilakos A, Cao H and Leung V C M, 2011, \emph{Body Area Networks: A Survey}. Mobile Networks and Applications \textbf{16}, 171.

\bibitem{UWB1d} Minasian R A, 2016, \emph{Ultra-Wideband and Adaptive Photonic Signal Processing of Microwave Signals}. IEEE J. Quantum Electronics, \textbf{52}, 1.

\bibitem{UWB2} Allen B, Dohler M, Okon E E, Malik W Q, Brown A K and Edwards D J, Eds, 2006, \emph{Ultra Wide Band Antennas and Propagation for Communications, Radars and Imaging}. Wiley, London.

\bibitem{UWB3} Wiesneck W, Adamiuk G and Sturm C, 2009, \emph{Basic properties and design principles of UWB antennas}. Proc. of the IEEE \textbf{97} 372.

\bibitem{UWB4} Adamiuk G, Zwick T and Wiesneck W, 2012, \emph{UWB antennas for communication systems}. Proc. of the IEEE \textbf{100} 2308.

\bibitem{UWB5} Ciccheti R, Miozzi E, Testa D, 2017, \emph{Wideband and UWB Antennas for Wireless Applications: A Comprenhensive Review}. International Journal of Antennas and Propagation, Article ID 2390808.


\bibitem{Rumsey} Rumsey V H, 1957, \emph{Frequency independent antennas}. IRE National Convention Record, \textbf{5} 114.

\bibitem{Mushiake1} Mushiake Y, 1948, \emph{The input impedance of a slit antenna}. Joint Convention Record of Tohoku Sections of IEE and IECE of Japan.

\bibitem{Nipones} Urade Y, Nakata Y, Nakanishi T and Kitano M, 2015, \emph{Frequency-independent response of self-complementary checkerboard screens}. Phys. Rev. Lett. \textbf{114} 237401.

\bibitem{Bateman1} Bateman H, 1908, \emph{The conformal transformations of a space of four dimensions and their application to geometrical optics}. Proc. Lond. Math. Soc. \textbf{7} 70.

 \bibitem{Cuni} Cunningham E, 1909, \emph{The principle of Relativity in electrodynamics and an extension thereof}. Proc. Lond. Math. Soc. \textbf{8} 77.

\bibitem{Bateman} Bateman H, 1910, \emph{The transformation of the electrodynamical equations}. Proc. Lond. Math. Soc. \textbf{8} 223.

\bibitem{Fulton} Fulton T, Rohrligh F and Witten L, 1962, \emph{Conformal invariance in physics}. Rev. Mod. Phys. \textbf{34} 442.

\bibitem{Conformal} Di Francesco P, Mathieu P and Senechal D, 1997, \emph{Conformal field theory}. Springer.

\bibitem{Kastrup} Kastrup HA, 2008, \emph{On the advancements of conformal transformations and their associated symmetries in geometry and theoretical physics}. Ann. Phys. (Berlin) \textbf{17} 631.

\bibitem{Osborn} Codirla C and Osborn H, 1997, \emph{Conformal invariance and electrodynamics}. Ann. Phys. \textbf{260} 91.

\bibitem{Isham} Isham C J, Salam A and Strathdee J, 1970 \emph{Broken Chiral and conformal symmetry in an effective-Lagrangian formalism}. Phys. Rev. \textbf{D2} 685.
\bibitem{Mushiake2} Mushiake Y, 1996, \emph{Self-Complementary Antennas: principle of self-complementarity for constant impedance}. Springer.

\bibitem{Booker} Booker H G, 1946, \emph{Slot aerials and their relation to complementary wire aerials (Babinet's principle)}. JIEE \textbf{IIIA}, \textbf{93} 620.
\bibitem{Deschamps} Deschamps G A, 1959, \emph{Impedance properties of complementary multiterminal planar structures}. IRE Trans. Ant. Prop. \textbf{AP-7} 371.

\bibitem{Collin} Collin R E, 1990, \emph{Field theory of guided waves}. IEEE Press, New York.

\bibitem{Balanis} Balanis C A, 2012, \emph{Advanced engineering electromagnetics}, Wiley.

\bibitem{Penrose1} Penrose R, 1965, \emph{Zero rest-mass fields including gravitation: asymptotic behaviour}. Proc. R. Soc. Lond. \textbf{A284} 159.

\bibitem{Penrose2} Penrose R, 1968, \emph{Structure of space-time}. Battelle Rencontres, 1967 lectures in mathematics and physics, edited by C. DeWitt and J. A. Wheeler (W. A. Benjamin, Inc, New York) 121.

\bibitem{ConformalAnt} Josefsson L and Persson P, 2006, \emph{Conformal array antenna theory and design}. IEEE Press.

\bibitem{H-E} Hawking S W and Ellis G F R, 1973, \emph{The large scale structure of spacetime}. Cambridge University Press.

\bibitem{UWB6} Morbidel L, Bulus Rossini L A and Costanzo Caso P A, 2017, \emph{Design of high return
loss logarithmic spiral antenna}. Microw. Opt. Technol. Lett. \textbf{59}, 2532.

\bibitem{UWB7} Seok H C, Jong K P, Sun K K and Jae Y P, 2004, \emph{A new ultra wide band antenna for UWB applications}. Microw. Opt. Technol. Lett. \textbf{40}, 399.
\bibitem{Power1} Elmansouri M A and Filipovic D S, 2012, \emph{Low-dispersion spiral antennas}. IEEE Trans. Antennas Propag. \textbf{60} 5522.

\bibitem{Power2} Huifen Huang and Zonglin Lv, 2014, \emph{A new spiral antenna with improved axial ratio and shorted arm lenght}. PIER C \textbf{46} 83.

\bibitem{Gustav} Gustafsson M, Sohl C and Kristensson G, 2007, \emph{Physical limitations of antennas of arbitrary shape}. Proc. R. Soc. Lond. \textbf{A463} 2589.
\end{thebibliography}
\end{document}